\documentstyle[12pt,psfig,epsf]{article}

\def\thefootnote{\fnsymbol{footnote}}

\begin{document}

\newcommand{\beq}{\begin{eqnarray}}	
\newcommand{\eeq}{\end{eqnarray}}	
\newcommand{\beqstar}{\begin{eqnarray*}}	
\newcommand{\eeqstar}{\end{eqnarray*}}	

\newcommand{\gsim}{ \mathop{}_{\textstyle \sim}^{\textstyle >} }
\newcommand{\lsim}{ \mathop{}_{\textstyle \sim}^{\textstyle <} }
\newcommand{\vev}[1]{ \left\langle {#1} \right\rangle }
\newcommand{\lsp}{ \left ( }
\newcommand{\rsp}{ \right ) }
\newcommand{\lmp}{ \left \{ }
\newcommand{\rmp}{ \right \} }
\newcommand{\llp}{ \left [ }
\newcommand{\rlp}{ \right ] }
\newcommand{\labs}{ \left | }
\newcommand{\rabs}{ \right | }

\newcommand{\K}{ {\rm K} }
\newcommand{\EV}{ {\rm eV} }
\newcommand{\KEV}{ {\rm keV} }
\newcommand{\MEV}{ {\rm MeV} }
\newcommand{\GEV}{ {\rm GeV} }
\newcommand{\TEV}{ {\rm TeV} }
\def\gtap{\raisebox{-.4ex}{\rlap{$\sim$}} \raisebox{.4ex}{$>$}}
\newcommand{\gggh}{${\rm G_{GUT}\times G_H}$ }

\begin{titlepage}
\begin{center}

\hfill    LBNL-39155\\
\hfill    UCB-96/35\\
\hfill    hep-ph/9607463\\
\hfill    July, 1996

\vskip .5in

{\Large \bf Non-unified gaugino masses in supersymmetric missing partner
models with hypercolor}\footnote
{This work was supported in part by the Director, Office of 
Energy Research, Office of High Energy and Nuclear Physics, Division of 
High Energy Physics of the U.S. Department of Energy under Contract 
DE-AC03-76SF00098 and in part by the National Science Foundation under 
grant PHY-95-14797.}

\vskip .5in

{\large Nima Arkani-Hamed$^{a,b}$, Hsin-Chia Cheng$^{a,b}$
and Takeo Moroi$^{a}$}

\vskip .5in

$^a${\it Theoretical Physics Group, 
Lawrence Berkeley National Laboratory\\
University of California, Berkeley, CA 94720, U.S.A.}

\vskip .2in

$^b${\it Department of Physics,
University of California, Berkeley, CA 94720, U.S.A.}

\end{center}

\vskip .5in

\begin{abstract}

The gaugino mass relations $m_3/{g_3^2} = m_2/{g_2^2} = m_1/{g_1^2}$ are 
considered to be  robust signals for supersymmetric grand unification.
In this letter, we point out that these relations may be significantly
modified in an interesting class of models which solve the doublet-triplet
splitting problem using a missing partner mechanism together with a strong
hypercolor gauge group. The observation of non-unified gaugino masses,
together with unified sfermion masses, provides a distinctive
signature for these models. 
 
\end{abstract}
\end{titlepage}
\renewcommand{\thepage}{\roman{page}}
\setcounter{page}{2}
\mbox{ }

\vskip 1in

\begin{center}
{\bf Disclaimer}
\end{center}

\vskip .2in

\begin{scriptsize}
\begin{quotation}
This document was prepared as an account of work sponsored by the United
States Government. While this document is believed to contain correct
information, neither the United States Government nor any agency
thereof, nor The Regents of the University of California, nor any of their
employees, makes any warranty, express or implied, or assumes any legal
liability or responsibility for the accuracy, completeness, or usefulness
of any information, apparatus, product, or process disclosed, or represents
that its use would not infringe privately owned rights.  Reference herein
to any specific commercial products process, or service by its trade name,
trademark, manufacturer, or otherwise, does not necessarily constitute or
imply its endorsement, recommendation, or favoring by the United States
Government or any agency thereof, or The Regents of the University of
California.  The views and opinions of authors expressed herein do not
necessarily state or reflect those of the United States Government or any
agency thereof of The Regents of the University of California and shall
not be used for advertising or product endorsement purposes.
\end{quotation}
\end{scriptsize}

\vskip 2in

\begin{center}
\begin{small}
{\it Lawrence Berkeley Laboratory is an equal opportunity employer.}
\end{small}
\end{center}
  
\newpage

\renewcommand{\thepage}{\arabic{page}}
\setcounter{page}{1}
\renewcommand{\thefootnote}{\arabic{footnote}}
\setcounter{footnote}{0}

Supersymmetric grand unified theory (SUSY-GUT)~\cite{SUSY_GUT} is
one of the most attractive candidates of new physics beyond the
standard model. It is supported by the precision measurements of the
gauge coupling constants of 
SU(3)$_{\rm C}$ $\times$ SU(2)$_{\rm L}$ $\times$
U(1)$_{\rm Y}$~\cite{gc_unification}.
However, it may be that the unification of gauge 
couplings is an accident. What further evidence can support the
existence of a SUSY-GUT? If supersymmetric particles are discovered, we may
hope to see signatures for unification in the superpartner spectrum.
In particular, the pattern of sfermion masses can probe the unification
of quark and lepton multiplets, while the gaugino masses can signal the
existence of a unified gauge group at the GUT scale.  Between the two, the 
weak scale scalar mass relations are more sensitive to 
the physics between the weak and 
GUT scales,  whereas the gaugino mass relations\footnote
{We neglect the higher loop effects of SU(3)$_{\rm C}$ $\times$
SU(2)$_{\rm L}$ $\times$ U(1)$_{\rm Y}$ gauge coupling constants,
as they are quite small.}
\beq
\frac{m_3}{g_3^2} : \frac{m_2}{g_2^2} : \frac{m_1}{g_1^2} =
\frac{m_{\rm GUT}}{g_{\rm GUT}^2} :
\frac{m_{\rm GUT}}{g_{\rm GUT}^2} :
\frac{m_{\rm GUT}}{g_{\rm GUT}^2}
= 1 : 1 : 1,
\label{gut-relation}
\eeq
where $g_3$, $g_2$, $g_1$ and $g_{\rm GUT}$ ($m_3$, $m_2$, $m_1$ and
$m_{\rm GUT}$) represent the gauge coupling constants (gaugino masses)
for SU(3)$_{\rm C}$, SU(2)$_{\rm L}$, U(1)$_{\rm Y}$, and G$_{\rm
GUT}$, respectively,\footnote
{We choose the GUT normalization for $g_1$, {\it i.e.}, 
$g_1=\sqrt{{5\over 3}}
g_{\rm Y}$.}
are a more robust prediction of gauge unification. However, these
simple relations (\ref{gut-relation}) may also 
arise in models with {\it no}
unified gauge group at high scales, as in some string theories with
dilaton-dominated SUSY breaking~\cite{NPB422-125}, or theories
with low energy dynamical SUSY breaking~\cite{DSB}.\footnote{In the case of
low energy dynamical SUSY breaking, the usual gaugino mass
relations only follow 
if the vector-like fields transmitting SUSY 
breaking to the ordinary sector form complete representations of SU(5),
which can perhaps be taken as indirect evidence for unification.}
Thus, we conclude that if the sfermion masses do not satisfy the GUT mass
relations, grand unification is by no means ruled out.
On the other hand,
verifying the gaugino mass relations, while extremely exciting,
would not suffice as a proof
for grand unification. 

However, what if the opposite happened?  Then there would be little
question as to quark-lepton unification, but what of gauge
unification? In this letter, we point out that non-unified gaugino masses
can in fact arise naturally in a very interesting class of grand
unified theories, providing a unique signature for these type of
models and a window into physics above the GUT scale.
 
Grand unified models suffer from a serious problem, the
``doublet-triplet splitting problem''.  With a unified gauge group
like SU(5) or SO(10), Higgs doublets are accompanied by color-triplet
Higgses. Higgs doublets are responsible for the electro-weak symmetry
breaking, and hence their masses are of the order of electro-weak
scale. On the other hand, the stability of the nucleon and/or
successful unification of the gauge coupling constants require the
colored Higgs masses to be of the order of the GUT scale, $M_{\rm
GUT}\sim 10^{16}{\rm GeV}$, which is much larger than the doublet
Higgs masses~\cite{hc_mass}.  In the minimal SUSY SU(5) model, this mass
hierarchy is obtained by an extreme fine tune among several parameters
in the superpotential. Many attempts have been made to solve this
problem~\cite{missing,DW,PGB}.

Recently, an interesting mechanism has been proposed to solve the
doublet-triplet splitting
problem~\cite{PLB344-211,PRD53-3913,9511431,9602439}. It is based
on an enlarged gauge group, G$_{\rm GUT}$ $\times$ G$_{\rm H}$, like
SU(5)$_{\rm GUT}$ $\times$ SU(3)$_{\rm H}$ ($\times$ U(1)$_{\rm H}$)
or SO(10)$_{\rm GUT}$ $\times$ SO(6)${\rm _H}$ (where the subscript "H" 
stands for hypercolor). A characteristic feature of these models is that
the SU(3)$_{\rm C}$ group is a diagonal subgroup of SU(3)$_{\rm GUT}$
and SU(3)$_{\rm H}$, where SU(3)$_{\rm GUT}$ $\in$ G$_{\rm GUT}$ and
SU(3)$_{\rm H}$ $\in$ G$_{\rm H}$, while SU(2)$_{\rm L}$ is embedded
only in G$_{\rm GUT}$. The doublet-triplet splitting 
problem is solved by the missing partner
mechanism~\cite{missing}, which in this case can work with smaller matter 
multiplet representations, guaranteeing a  perturbative picture of G$_{\rm
GUT}$ up to the Planck scale.

Even though the unified gauge group is not simple, unification of the
gauge coupling constants of SU(3)$_{\rm C}$ $\times$ SU(2)$_{\rm L}$
$\times$ U(1)$_{\rm Y}$ is not spoiled if the gauge couplings
of G$_{\rm H}$ are large enough. In this case, the corrections to the gauge
couplings of the low energy gauge group can be smaller than what can 
be distinguished by precision tests of the gauge coupling constants.
However, we will show that the gaugino
masses may deviate from the usual GUT relation (1), while
scalar masses should still be unified at the GUT scale if they
belong to the same multiplet of the GUT group. 

Let us first review the \gggh model. To make our points clear, we 
will concentrate on the model based on the gauge group SU(5)$_{\rm GUT}$
$\times$ SU(3)$_{\rm H}$ $\times$ U(1)$_{\rm H}$ given in
Ref.~\cite{PRD53-3913}. The generalization to other models is 
straightforward.

In the ${\rm SU(5)_{GUT} \times SU(3)_{H} \times U(1)_{H}}$ model,
the particle content which is responsible for the breaking of 
\gggh group
consists of the following chiral supermultiplets:
$Q({\bf 5^*}, {\bf 3}, 1)$, $\bar{Q}({\bf 5}, {\bf 3^*}, -1)$, $q({\bf
1}, {\bf 3}, 1)$, $\bar{q}({\bf 1}, {\bf 3^*}, -1)$, $\Sigma ({\bf
24}, {\bf 1}, 0)$, $H({\bf 5}, {\bf 1}, 0)$, and $\bar{H}({\bf 5^*},
{\bf 1}, 0)$, where the numbers in brackets denote the transformation
properties under SU(5)$_{\rm GUT}$, SU(3)$_{\rm H}$ and 
U(1)$_{\rm H}$, respectively.
The superpotential of the model is given by 
\beq
W = \bar{Q}^\alpha_A (m_Q \delta^A_B + \lambda \Sigma^A_B) Q^B_\alpha
+ \frac{1}{2} m_\Sigma {\rm tr}(\Sigma^2)
+ h H_A Q^A_\alpha \bar{q}^\alpha 
+ h' \bar{H}^A \bar{Q}_A^\alpha q_\alpha,
\label{W}
\eeq
where $\lambda$, $h$ and $h'$ are dimensionless coupling constants,
while $m_Q$ and $m_\Sigma$ are mass parameters of the 
order of the GUT scale
$M_{\rm GUT}\sim 10^{16}{\rm GeV}$. Minimizing the potential,
we find that there is a vacuum in which the 
scalar components of $Q$, $\bar{Q}$, and $\Sigma$
have the following vacuum expectation values (VEVs),
\beq
&&\langle Q \rangle = \langle \bar{Q}^{\rm T} \rangle = 
\left(
\begin{array}{ccccc}
0&0&v&0&0 \\ 0&0&0&v&0 \\ 0&0&0&0&v
\end{array}
\right),
\label{vev_Q}\\
&&\langle\Sigma\rangle = \frac{m_Q}{2\lambda}
{\rm diag}(3,3,-2,-2,-2),
\label{vev_Sigma}
\eeq
with $v^2=5m_Qm_\Sigma /\lambda^2$, while other fields do not acquire a VEV.

In the vacuum given in eqs.(\ref{vev_Q}) and (\ref{vev_Sigma}), the
gauge symmetry SU(5)$_{\rm GUT}$ $\times$ SU(3)$_{\rm H}$ $\times$
U(1)$_{\rm H}$ is broken down to the standard model gauge group
SU(3)$_{\rm C}$ $\times$ SU(2)$_{\rm L}$ $\times$
U(1)$_{\rm Y}$ at the GUT scale. As we can see from the last two terms in
eq.(\ref{W}),
the colored Higgses, $H_I$ and $\bar{H}^I$ ($I$ = 3 -- 5), get large masses
from the $Q,\bar{Q}$ VEVs  
by marrying with $q^\alpha$ and $\bar{q}_\alpha$, 
while doublet Higgses, $H_i$ and
$\bar{H}^i$ ($i$ = 1, 2), remain massless.  Thus, the doublet-triplet
splitting is naturally achieved in this model due to the missing
partner mechanism.

An important feature of the \gggh model is that the low energy SU(3)$_{\rm C}$
symmetry is the diagonal subgroup of SU(3)$_{\rm GUT}$ $\times$
SU(3)$_{\rm H}$, where SU(3)$_{\rm GUT}$ is embedded in SU(5)$_{\rm
GUT}$. The gauge field of SU(3)$_{\rm C}$, $G_\mu$, is
given by a linear combination of the gauge field of SU(5)$_{\rm
GUT}$, $A_{{\rm GUT},\mu}$, and of SU(3)$_{\rm H}$, $A_{\rm
H3,\mu}$, as
\beq
G_\mu = \frac{1}{\sqrt{g_{\rm H3}^2+g_{\rm GUT}^2}}
(g_{\rm H3} A_{\rm GUT,\mu}
+ g_{\rm GUT} A_{\rm H3,\mu}),
\label{gluon}
\eeq
where $g_{\rm GUT}$ and $g_{\rm H3}$ are the gauge
coupling constants for SU(5)$_{\rm GUT}$ and SU(3)$_{\rm H}$ at the GUT
scale, and the gauge coupling  $g_3$ of SU(3)$_{\rm C}$ is given by, (in
the following equations (6) -- (11), all relations are understood to 
hold at the GUT scale) 
\beq
g_3^2 = 
\frac{g_{\rm H3}^2g_{\rm GUT}^2}{g_{\rm H3}^2+g_{\rm GUT}^2},
\label{gH3}
\eeq
or in terms of $\alpha(={g^2/4\pi})$,
\beq
{1\over \alpha_3}= {1\over \alpha_{\rm GUT}} + {1\over \alpha_{\rm H3}}.
\label{alphaH3}
\eeq
Similarly, U(1)$_{\rm Y}$ is  a subgroup of
U(1)$_{\rm GUT}$ $\times$ U(1)$_{\rm H}$, and its gauge field
$B_\mu$ and gauge coupling constant $g_1$ are given by
\beq
B_\mu&=&\frac{1}{\sqrt{15g_{\rm H1}^2+g_{\rm GUT}^2}}
(\sqrt{15}g_{\rm H1} A_{\rm GUT,\mu}
+ g_{\rm GUT} A_{\rm H1,\mu}),
\\
g_1^2&=&
\frac{15g_{\rm H1}^2g_{\rm GUT}^2}{15g_{\rm H1}^2+g_{\rm GUT}^2}.
\eeq
On the other hand, SU(2)$_{\rm L}$ is embedded only in SU(5)$_{\rm
GUT}$, and hence its gauge field $W_\mu$ and gauge
coupling constant $g_2$ are given by
\beq
W_\mu&=&A_{{\rm GUT},\mu},
\\
g_2^2&=&g_{\rm GUT}^2.
\label{g2}
\eeq
The important point is that $g_1(M_{\rm GUT})\simeq g_2(M_{\rm GUT})
\simeq g_3(M_{\rm GUT})$ if $g_{\rm H3}(M_{\rm
GUT})$, $g_{\rm H1}(M_{\rm GUT})$ $\gg$ $g_{\rm GUT}(M_{\rm GUT})$.
Thus, the gauge coupling unification is not spoiled if the gauge
coupling constants of G$_{\rm H}$ are large enough. 
As a result, it is
very difficult to distinguish \gggh model with the ordinary
GUTs by the precise measurements of the gauge coupling
constants.
In fact, recent analyses show that the predicted strong coupling
constant from SUSY-GUT without including threshold corrections,
$\alpha_s(M_Z)=0.130$ \cite{LP,BP}, is a little bit higher than the world
averaged experimental value, $\alpha_s(M_Z)= 0.118\pm 0.003$ \cite{alphas}.
We can see from eqs.(\ref{gH3}) and (\ref{alphaH3}) that 
the correction from the hypercolor
gauge coupling reduces the SU(3)$_{\rm C}$ gauge coupling, and hence shifts
the prediction in the right direction to be consistent with the 
experimental value. The correction from the U(1)$_{\rm H}$ coupling
also moves the prediction in the right direction by changing the unification
scale. 
%To see this and also put the lower
%bound on the SU(3)$_{\rm H}$ and U(1)$_{\rm H}$ couplings, it is 
%convenient to express the corrections from the G$_{rm H}$ couplings
%on the Weinberg angle. 
Let the shift of the inverse couplings of
SU(3)$_{\rm C}$ and U(1)$_{\rm Y}$ at the GUT scale due to G$_{\rm H}$
couplings be $\delta_3$ and $\delta_1$, 
\beq
\delta_3 = {1\over \alpha_3(M_G)}-{1\over \alpha_{\rm GUT}(M_G)} 
={1\over \alpha_{\rm 3H}}, \;\;\;
\delta_1 = {1\over \alpha_1(M_G)}-{1\over \alpha_{\rm GUT}(M_G)}
={1\over 15 \alpha_{\rm 1H}}.
\eeq
A simple calculation using one-loop renormalization group equations
(RGEs) gives the shift of the inverse of the
predicted strong coupling constant due to $\delta_3$ and
$\delta_1$ as
\beq
\Delta \left( {1\over \alpha_s} \right)
= \delta_3 + {5\over 7} \delta_1.
\eeq
So, the prediction of $\alpha_s (M_Z)$ in this model can
be written as 
\beq
\alpha_s(M_Z) \simeq 0.130 - {0.014\over \alpha_{\rm 3H}}
-{0.010\over 15 \alpha_{\rm 1H}} + \Delta_{\alpha_s},
\eeq
where the first term is the ordinary SUSY-GUT prediction, and
the last term represents the extra threshold corrections. 
Typically it is found that $|\Delta_{\alpha_s}| < 0.01$ \cite{LP}.
Then 
in order to
be consistent with the experimental value, we require
$\alpha_{\rm 3H} \gtap 0.6$
(for $\delta_1=0$) and $\alpha_{\rm 1H} \gtap 0.03$ (for $\delta_3=0$).

Now, we are in a  position to discuss the gaugino masses in the \gggh
model. The gaugino masses originate in the soft SUSY breaking terms,
whose origin is related to the mechanism of the SUSY breaking.
In the main part of this letter, we assume a hidden-sector SUSY breaking
scenario, in which SUSY breaking is mediated by supergravity. 
The gauginos have the following mass terms above the GUT scale,
\beq
{\cal L} = -\frac{1}{2} m_{\rm GUT}\lambda_{\rm GUT}\lambda_{\rm GUT}
-\frac{1}{2} m_{\rm H3}\lambda_{\rm H3}\lambda_{\rm H3}
-\frac{1}{2} m_{\rm H1}\lambda_{\rm H1}\lambda_{\rm H1}
+ h.c.,
\label{L_gaugino}
\eeq
where $\lambda_{\rm GUT}$, $\lambda_{\rm H3}$ and $\lambda_{\rm H1}$
($m_{\rm GUT}$, $m_{\rm H3}$ and $m_{\rm H1}$) are gauginos (gaugino
masses) for SU(5)$_{\rm GUT}$, SU(3)$_{\rm H}$ and U(1)$_{\rm H}$
gauge groups, respectively. Heavy particles decouple at the GUT scale,
and below the GUT scale, we have the SUSY standard model as a
low energy effective theory. In particular, gauginos for the standard
model gauge group SU(3)$_{\rm C}$, SU(2)$_{\rm L}$ and U(1)$_{\rm Y}$
(which we denote $\tilde{G}$, $\tilde{W}$ and $\tilde{B}$), are given
by, (eqs.(16) -- (22) are understood to hold at the GUT scale)
\beq
\tilde{G} &=& \frac{1}{\sqrt{g_{\rm H3}^2+g_{\rm GUT}^2}}
(g_{\rm H3} \lambda_{\rm GUT} + g_{\rm GUT} \lambda_{\rm H3}),
\label{gluino} \\
\tilde{W} &=& \lambda_{\rm GUT},
\label{wino} \\
\tilde{B} &=& \frac{1}{\sqrt{15g_{\rm H1}^2+g_{\rm GUT}^2}}
(\sqrt{15}g_{\rm H1} \lambda_{\rm GUT} + g_{\rm GUT} \lambda_{\rm
H1}).
\label{bino}
\eeq
Substituting eqs.(\ref{gluino}) --
(\ref{bino}) into eq.(\ref{L_gaugino}), we obtain masses for
$\tilde{G}$, $\tilde{W}$ and $\tilde{B}$ as
\beq
m_3 &=& g_3^2\left(
\frac{m_{\rm H3}}{g_{\rm H3}^2} + \frac{m_{\rm GUT}}{g_{\rm GUT}^2} \right), 
\label{m3} \\ 
m_2 &=& m_{\rm GUT}, 
\label{m2} \\
m_1 &=& g_1^2 \left(\frac{m_{\rm H1}}{15g_{\rm H1}^2} + \frac{m_{\rm
GUT}}{g_{\rm GUT}^2} \right). 
\label{m1} 
\eeq 
We can see that the GUT relation on the gaugino masses
(\ref{gut-relation}) is modified
\beq
\frac{m_3}{g_3^2} : \frac{m_2}{g_2^2} : \frac{m_1}{g_1^2} =
\left(
\frac{m_{\rm GUT}}{g_{\rm GUT}^2} + \frac{m_{\rm H3}}{g_{\rm H3}^2}
\right) :
\frac{m_{\rm GUT}}{g_{\rm GUT}^2} :
\left(
\frac{m_{\rm GUT}}{g_{\rm GUT}^2} + \frac{m_{\rm H1}}{15g_{\rm H1}^2}
\right).
\label{new-relation}
\eeq
The above relations receive negligible modification in running from
the GUT to the weak scale.  Thus, if the ratio $m_{\rm H3}/g_{\rm
H3}^2$ or $m_{\rm H1}/g_{\rm H1}^2$ at the GUT scale is
comparable to $m_{\rm GUT}/g_{\rm GUT}^2$, significant deviations from
the usual GUT relations (\ref{gut-relation}) can be observed when
gaugino masses are measured.  Notice that the combinations $m_{\rm
H3}/g_{\rm H3}^2$ and $m_{\rm H1}/g_{\rm H1}^2$ are renormalization
group invariants at the one loop level. So, in contrast with the gauge
coupling, at one loop the corrections to the gaugino mass relations do
not diminish as the hyper gauge couplings become large.  On the other
hand, in this model, squarks and sleptons are contained in $({\bf
5^*}, {\bf 1}, 0)$ or $({\bf 10}, {\bf 1}, 0)$ representation of the
unified gauge group, as in the ordinary GUT. Thus, sfermion mass
unification is still expected. These facts suggest that the mass
spectroscopy of the superparticles can give us a signal for these kind
of models.

Now, let us discuss the magnitude of $m_{\rm H}/g_{\rm H}^2$. In the
hidden sector SUSY breaking scenario, gaugino masses are usually given
in the form
\beq
{\cal L} = \sum_{\rm G} \int d^2\theta
\frac{k_{\rm G}}{M_{\rm PL}} S W^{\rm G} W^{\rm G} + h.c.,
\label{sww}
\eeq
where $S$ denotes the chiral multiplet which is responsible for the
SUSY breaking, $M_{\rm PL}\simeq 2.4\times 10^{18}$GeV is the reduced Planck
scale, $k_{\rm G}$ denote the coupling constants, and $W^{\rm
G}$ is the superfield for the gauge multiplet. (Here, G indicates
gauge group.) Then, when SUSY is broken ($F_S \equiv \langle\int d^2\theta
S\rangle \neq 0$), we will get gaugino masses of order
$k_{\rm G}F_S/M_{\rm PL}$. In general, we do not expect that
$k_{\rm H3}$
and $k_{\rm H1}$ are much smaller than $k_{\rm GUT}$, and hence the deviation 
from the GUT relation is expected to be non-negligible.
To make a more 
definite statement we have to
make assumptions about how SUSY is broken. For example, 
in the superstring inspired model with 
dilaton-dominated SUSY breaking~\cite{NPB422-125}, the 
combination $m_{\rm G}/g_{\rm G}^2$ is universal for all the gauge
groups G at the string scale. Then, large corrections to the GUT gaugino 
mass 
relations are expected. In particular, the gaugino masses for
SU(3)$_{\rm C}$ and SU(2)$_{\rm L}$ obey
$m_3/g_3^2:m_2/g_2^2=2:1$, neglecting the higher order 
corrections.\footnote
{For U(1)$_{\rm H}$, we do not know the normalization of the 
charges of the chiral multiplets, and hence we 
cannot give definite prediction on U(1)$_{\rm Y}$.}
In general, the ratio, 
\beq
R_{3/2}\equiv {m_3/g_3^2 \over m_2/g_2^2},
\eeq
is related to the boundary condition, 
\beq
R_{\rm H/G}\equiv 
\left. {m_{\rm H3}/g_{\rm H3}^2 \over m_{\rm GUT}/ g_{\rm GUT}^2}
\right|_{\mu =M_{\rm PL}}, 
\eeq
as $R_{3/2}=R_{\rm H/G} +1$ in the one
loop approximation (with $\mu$ being the renormalization point).

So far, we considered only one loop RGEs.
However, as we discussed before,
gauge coupling constants for ${\rm G_H}$ have to be large at the
GUT scale, and hence the results based on the one loop RGEs may not be
a good approximation. In fact, the ratio of the gaugino mass to the
gauge coupling constant squared receives higher order corrections and
does not remain constant. Therefore, $R_{3/2}$ will depend explicitly
on $g_{\rm H3}(M_G)$ as well as on $R_{\rm H/G}$
if we take into
account the higher loop effects due to the large coupling $g_{\rm H3}$.

In order to demonstrate the effect of the higher order terms, we use
two loop RGEs to evolve the ratios between the GUT scale and the 
Planck scale. The two loop RGEs for gauge couplings and gaugino masses 
are given in \cite{PRL72-25}. 
We fix $\alpha_{\rm GUT}(= g^2_{\rm GUT}/ 4\pi)= 1/25$ at the GUT scale,
and numerically evaluate $R_{3/2}$ as a function of $\alpha_{\rm H3}
(M_{\rm GUT})$ for different initial values of $R_{\rm H/G}$ at
$M_{\rm PL}$. The result is shown in Fig.~1 for $0.5<\alpha_{\rm H3}
(M_{\rm GUT}) < 2$. We can see that $R_{3/2}$ is close to the 
one loop value, $R_{\rm H/G}+1$, for smaller $\alpha_{\rm H3}$ and
$R_{\rm H/G}$, but deviates significantly from the one loop result for larger 
$\alpha_{\rm H3}$ and $R_{\rm H/G}$. Notice that the 
apparent blow up of $R_{3/2}$ 
in the case with $R_{\rm H/G} = 2$ is due to $m_{\rm GUT}$ being scaled to
zero in the course of running. Below $M_{\rm GUT}$, $R_{3/2}$
stays approximately constant since there is no large coupling
to make  higher loop contributions important. For very
large $\alpha_{\rm H3}$, the perturbative calculation should
break down and the results based on the two loop calculation are not 
reliable. In that case, we have no control on $m_{\rm H3}/g_{\rm H3}^2$
near the GUT scale. However, based on our results for moderate large
$\alpha_{\rm H3}$, we do not expect that $m_{\rm H3}/g_{\rm H3}^2$
quickly goes to zero for finite values of $\alpha_{\rm H3}$.

Some comments are in order. First of all, we would like to discuss
the models based on the gauge group other than SU(5)$_{\rm GUT}$
$\times$ SU(3)$_{\rm H}$ $\times$ U(1)$_{\rm H}$, {\it i.e.} models
based on SU(5)$_{\rm GUT}$ $\times$ SU(3)$_{\rm H}$~\cite{9511431} or
SO(10)$_{\rm GUT}$ $\times$ SO(6)${\rm _H}$~\cite{9602439}. In those
cases, SU(3)$_{\rm C}$ is a diagonal subgroup of SU(3)$_{\rm GUT}$ and
SU(3)$_{\rm H}$ as in the previous case, while SU(2)$_{\rm L}$ and
U(1)$_{\rm Y}$ are embedded only in G$_{\rm GUT}$. Then, the gaugino
masses  obey the relation (\ref{new-relation}) with
$m_{\rm H1}=0$, {\it i.e.}, the gaugino masses for SU(2)$_{\rm L}$ and
U(1)$_{\rm Y}$ obey the usual GUT relation, while that for SU(3)$_{\rm
C}$ does not. This kind of signal, together with the unifications of
the sfermion masses, will give us an information on the 
structure of the \gggh
model.

So far, we have concentrated on the hidden sector SUSY breaking
scenario. There is another interesting scenario where supersymmetry
is broken dynamically at a low energy scale, then mediated to the 
observable sector by a messenger sector~\cite{DSB}. To preserve gauge 
coupling
unification, the messenger sector should fill out complete
multiplets under ordinary SU(5). In this case, 
the GUT relation (1) is not affected
even in the G$_{\rm GUT}\times$G$_{\rm H}$ model. 

In summary, we have investigated the gaugino masses in supersymmetric
unified models  based on an enlarged gauge group like SU(5)$_{\rm
GUT}$ $\times$ SU(3)$_{\rm H}$ ($\times$ U(1)$_{\rm H}$) or
SO(10)$_{\rm GUT}$ $\times$ SO(6)${\rm _H}$, in which the
doublet-triplet splitting problem can be solved naturally.  In these
models, the GUT relations on the gaugino masses can be broken completely or
partially, while we can still hope that the unification of the sfermion
masses is unaffected. Therefore, by the accurate spectroscopy of the
superparticles, we may have a window into physics at and beyond the GUT 
scale.

The authors would like to thank L.J. Hall and H. Murayama for useful
discussions. This work was supported in part by the Director, Office
of Energy Research, Office of High Energy and Nuclear Physics,
Division of High Energy Physics of the U.S. Department of Energy under
Contract DE-AC03-76SF00098 and in part by the National Science
Foundation under grant PHY-95-14797, and the work of N.A.-H. is 
supported by NSERC.

%%%%%%%%%%%%%%%%%%%%%%%%%%%%%%%%%%%%%%%%%%%%%%%%%%%%%%%%%%%%%%%
%
% NEW COMMANDS FOR THE BIBLIOGRAPHY
%
%%%%%%%%%%%%%%%%%%%%%%%%%%%%%%%%%%%%%%%%%%%%%%%%%%%%%%%%%%%%%%%
\newcommand{\Journal}[4]{{\sl #1} {\bf #2} {(#3)} {#4}}
\newcommand{\APJ}{Ap. J.}
\newcommand{\CJP}{Can. J. Phys.}
\newcommand{\NC}{Nuovo Cimento}
\newcommand{\NP}{Nucl. Phys.}
\newcommand{\MPL}{Mod. Phys. Lett.}
\newcommand{\PL}{Phys. Lett.}
\newcommand{\PR}{Phys. Rev.}
\newcommand{\PRep}{Phys. Rep.}
\newcommand{\PRL}{Phys. Rev. Lett.}
\newcommand{\PTP}{Prog. Theor. Phys.}
\newcommand{\SJNP}{Sov. J. Nucl. Phys.}
\newcommand{\ZP}{Z. Phys.}
%%%%%%%%%%%%%%%%%%%%%%%%%%%%%%%%%%%%%%%%%%%%%%%%%%%%%%%%%%%%%%%

\begin{figure}[p]
\epsfxsize=13cm
\centerline{\epsfbox{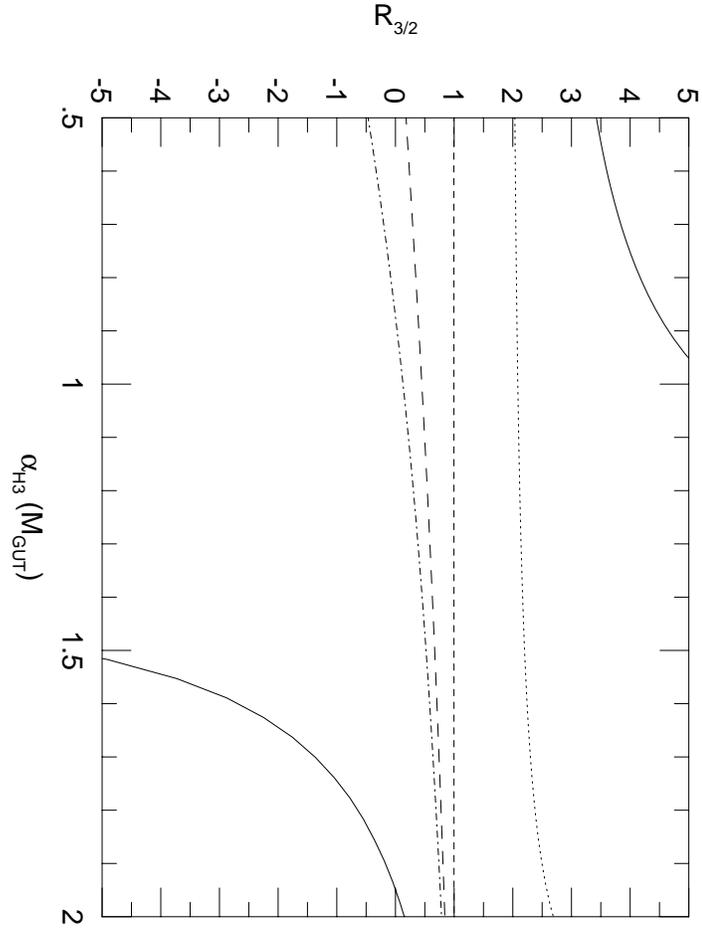}}
\vskip .5in
\caption{The ratio R$_{3/2}$ as a function of 
$\alpha_{\rm H3}(M_{\rm GUT})$ for different values of the ratio
R$_{\rm H/G}$=2 (solid), 1 (dotted), 0 (short dashed), $-1$ 
(long dashed) and $-2$ (dot-dashed).}
\end{figure}

\end{document}